\documentclass{mn2e}
\input{epsf}

\title[Radial velocity moments of dark matter haloes]
{Radial velocity moments of dark matter haloes}
\author[R. Wojtak et al.]{Rados{\l}aw Wojtak$^{1}$, Ewa L. {\L}okas$^{2}$,
Stefan Gottl\"ober$^{3}$
and Gary A. Mamon$^{4,5}$
\\   \\
$^1$Astronomical Observatory, Jagiellonian University, Orla 171, 30-244 Cracow,
Poland, {\rm radek\b{\ }wojtak@o2.pl} \\
$^2$Nicolaus Copernicus Astronomical Center, Bartycka 18, 00-716 Warsaw, Poland,
{\rm lokas@camk.edu.pl}\\
$^3$Astrophysikalisches Institut Potsdam, An der Sternwarte 16, 14482 Potsdam, Germany,
{\rm sgottloeber@aip.de} \\
$^4$Institut d'Astrophysique de Paris (CNRS UMR 7095), 98 bis Bd Arago, F-75014 Paris,
France, {\rm gam@iap.fr} \\
$^5$GEPI (CNRS UMR 8111), Observatoire de Paris, F-92195 Meudon, France}

\begin{document}

\maketitle

\vspace{-0.2in}

\begin{abstract}

Using cosmological $N$-body simulations we study the radial velocity distribution
in dark matter haloes focusing on the lowest-order even moments, dispersion and kurtosis.
We determine the properties of ten massive haloes in the simulation box
approximating their density distribution by the NFW formula characterized by the virial
mass and concentration. We also calculate the velocity anisotropy parameter of the haloes
and find it mildly radial and increasing with distance from the halo centre.
The radial velocity dispersion of the haloes shows a characteristic profile with a
maximum, while the radial kurtosis profile decreases with distance
starting from a value close to Gaussian near the centre. We therefore confirm
that dark matter haloes possess intrinsically non-Gaussian, flat-topped velocity distributions.
We find that the radial velocity moments of the simulated haloes
are quite well reproduced by the solutions of the Jeans equations obtained for
the halo parameters with the anisotropy measured in the simulations.
We also study the radial velocity moments
for a composite cluster made of ten haloes out to ten virial radii. In this region
the velocity dispersion decreases systematically to reach
the value of the background, while kurtosis increases from below to above the Gaussian value of
3 signifying a transition from a flat-topped to a strongly peaked velocity distribution
with respect
to the Gaussian, which can be interpreted as the dominance of ordered flow with a small
dispersion. We illustrate the transition by showing explicitly the velocity distribution
of the composite cluster in a few radial bins.

\end{abstract}

\begin{keywords}
methods: $N$-body simulations -- methods: analytical -- galaxies: clusters: general
-- galaxies: kinematics and dynamics -- cosmology: dark matter
\end{keywords}

\vspace{-0.25in}

\section{Introduction}

The density distribution of bound structures has been a subject of vigorous research in
recent years. An NFW (Navarro, Frenk \& White 1997)
density profile was established as a universal formula describing
dark matter haloes in large mass range arising in cosmological $N$-body simulations
(for recent refinements see Navarro et al. 2004; Diemand, Moore \& Stadel 2004a;
Tasitsiomi et al. 2004; Merritt et al. 2005).
Much less attention
has been devoted to the velocity distribution in bound structures, although it also
carries substantial amount of information about the formation and evolution of structure
in the Universe. The moments of the velocity distribution in particular proved
extremely useful in studies of mass distribution in galaxies and clusters
(Kronawitter et al. 2000; van der Marel et al. 2000; {\L}okas \& Mamon 2003).

Recently a few groups studied the velocity distribution of simulated haloes
showing, for example, that it departs significantly from a Gaussian distribution expected
for virialized objects (Kazantzidis et al. 2004; Sanchis, {\L}okas \& Mamon 2004;
Diemand Moore \& Stadel 2004b).
Such behaviour, manifesting itself in a form of kurtosis value lower than 3,
was also seen in real objects like the Coma cluster ({\L}okas \& Mamon 2003) or the Draco
dwarf ({\L}okas, Mamon \& Prada 2004).
There have been few attempts to provide a theoretical explanation for departures from
Gaussianity; it turns out however that weakly non-Gaussian velocity distributions arise
naturally in Jeans theory of equilibrium structures (Merrifield \& Kent
1990; {\L}okas 2002; Kazantzidis et al. 2004). Hansen et al. (2004) have shown that the
flattened velocity distribution can also be interpreted in terms of Tsallis statistics.

In this Letter we study the radial velocity distribution of dark matter haloes resulting
from cosmological $N$-body simulations focusing on the most useful even moments,
the dispersion and kurtosis. We show that they are quite well reproduced by the solutions
of the Jeans equations. We also look at the behaviour of the moments outside the virial radius
and the overall velocity distribution.

\vspace{-0.25in}

\section{The simulated dark matter haloes}

For this study we run a cosmological dark matter simulation
within a box of size 150 $h^{-1}$ Mpc assuming
the concordance cosmological model ($\Lambda$CDM) with
parameters $\Omega_M=0.3$, $\Omega_{\Lambda}=0.7$, $h=0.7$ and $\sigma_8=0.9$.
We have used a new MPI (Message-Passing Interface)
version of the original ART (Adaptive Refinement
Tree) code (Kravtsov et al. 1997). The ART code achieves high spatial
resolution by refining the underlying uniform grid in all high-density
regions with an automated refinement/derefinement algorithm. In the
new MPI version of the code the whole simulation box is subdivided
into cuboids which contain approximately the same numerical tasks to
achieve load balance (not necessarily the same number of particles).
Each cuboid is handled by one node of the computer using internal
OpenMP parallelization. At the given node the remaining box is evolved
using more massive particles as described for the multi-mass version
of the ART code (Klypin et al. 2001). This procedure minimizes the
communication between nodes. After each basic integration step the
borders of the cuboids can be moved if load balance changes.
The simulation used $256^3$ particles, therefore it achieved a mass
resolution of $1.7 \times 10^{10} h^{-1} M_{\sun}$. The basic grid of the
simulation was 256, the maximum refinement level was 6. Therefore, the
force resolution (2 cells) reached $18 h^{-1}$ kpc.

The halos have been identified using the hierarchical friends-of-friends (HFOF) algorithm
(Klypin et al. 1999) using different linking lengths. The linking length of $b=0.17$
extracts haloes with masses close to the virial mass for our cosmological model.
(We define the virial mass and radius as those with mean density $\Delta_c=101.9$
times the critical density, see {\L}okas \& Hoffman 2001.)
However, the FOF method is known to connect particles along thin
bridges which are likely to be broken once the linking parameter is changed. This also
means that the shape of structure formed by linked particles differs strongly from spherical
and the centre of a halo does not necessarily coincide with the
region of maximum density.

In order to find points of maximum density we cut out the particles out to a virial radius
(assuming that the FOF mass is equal to the virial mass) and calculate the centres of
mass in spheres of decreasing radius, each time making the centre of mass the new halo
centre. The centres determined in this way agree very well with centres found with much
smaller linking lengths which pick up regions of higher density.
We then determine the new masses and virial radii of
the haloes by summing up masses of all particles found inside the virial radius defined as
before but now measured with respect to the new centre. Choosing the haloes for the analysis
we have generally taken the most massive objects found in the simulation box. We have,
however, rejected those undergoing a major merger with two or more subhaloes of comparable
mass which would significantly depart from equilibrium. Within the virial radius,
the haloes typically have a few times $10^4$ particles. The virial masses and radii
of the haloes, $M_v$ and $r_v$, are listed in Table~\ref{properties}.

\begin{table}
\caption{Properties of the simulated haloes.}
\label{properties}
\begin{center}
\begin{tabular}{cccccc}
\hline
Halo & $M_v$ & $r_v$ & $c$ \\
	& [$10^{14} M_{\sun}$] & [Mpc] & \\
\hline
1    & 7.49  & 2.33  & 5.968    \\
2    & 7.04  & 2.29 & 10.098    \\
3    & 6.83  & 2.26  & 5.360    \\
4    & 6.49  & 2.23  & 8.301    \\
5    & 5.87  & 2.15  & 6.285    \\
6    & 5.35  & 2.08  & 10.320   \\
7    & 4.82  & 2.02  & 8.577    \\
8    & 4.25  & 1.93  & 3.552    \\
9    & 3.86  & 1.87  & 7.607    \\
10   & 3.54  & 1.82  & 7.424    \\ \hline
\end{tabular}
\end{center}
\end{table}

For each of the ten haloes we determined its density profile taking averages
in radial bins of equal logarithmic length up to the virial radius.
The data points obtained in this way were assigned errors estimated as Poisson
fluctuations. The measured density profiles are well
approximated by the NFW formula (Navarro et al. 1997)
$\varrho(s)/\varrho_{c,0} = \Delta_c \,c^2 g(c)/[3 \,s\,(1+ c s)^2]$
where $s=r/r_v$, $\varrho_{c,0}$ is the present critical density,
$\Delta_c=101.9$ is the characteristic density parameter,
$c$ is the concentration parameter and $g(c) = [\ln (1+c) -
c/(1+c)]^{-1}$. Fitting this formula to our measured density profiles we determined
the concentration parameters of the haloes which are listed in Table~\ref{properties}.
Our estimated concentrations are consistent with the dependence of $c$ on mass
inferred from $N$-body simulations by Bullock et al. (2001), also run with a $\Lambda$CDM
cosmology.

In the following, all velocities will be calculated with respect to the mean velocity
inside the virial radius of a given halo. For each halo we measure the mean radial
velocity (with the Hubble flow added) of dark matter particles enclosed in shells of thickness $0.1
r_v$ centered at $0.05 r_v$, $0.15 r_v$ etc. assuming a convention that the negative sign
indicates infall motion towards the centre of the halo and normalizing to
$V_v = (G M_v/r_v)^{1/2}$, the circular velocity at $r_v$. For a fully virialized object
the mean radial velocity should be zero. We find that the measured values are
consistent on average with zero inside $r_v$ for most haloes, although some radial
variations caused by internal streaming motions are present. They do not exceed $0.4 V_v$
however, indicating a rather relaxed state of the objects.

\begin{figure}
\begin{center}
    \leavevmode
    \epsfxsize=8cm
    \epsfbox[55 55 270 250]{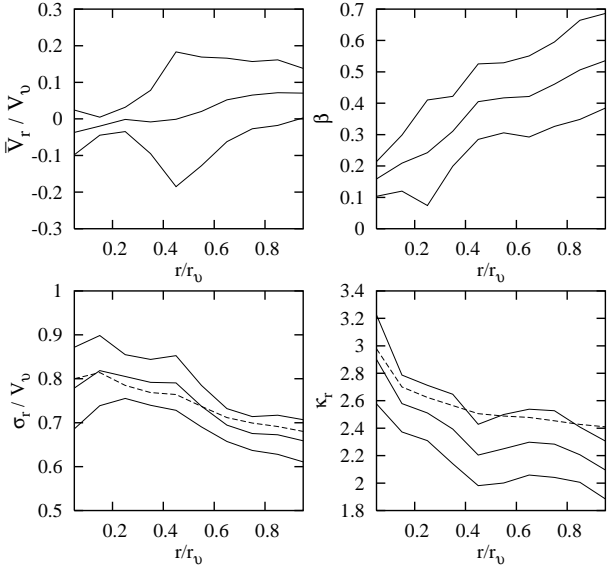}
\end{center}
    \caption{The mean radial velocity in units of $V_v$, the circular velocity at
	$r_v$ (upper left panel), the radial profile of the anisotropy
	parameter $\beta$ (upper right panel) and the radial moments:
	velocity dispersion $\sigma_r$ in units of $V_v$ (lower left panel) and
	kurtosis $\kappa_r$ (lower right panel)
	inside $r_v$ averaged over ten
	haloes listed in Table~\ref{properties}. Thicker solid lines show the mean and
	the thinner solid lines the dispersion in the measurements.
	Dashed lines give predictions from the Jeans equations for
	$c=7.4$ and a variable $\beta$ changing as shown in the upper right panel.}
\label{comparison}
\vspace{-0.15in}
\end{figure}

We then average our measurements calculating the mean and dispersion of the measurements for
the ten haloes in the corresponding bins. The results are plotted in the upper left panel of
Fig.~\ref{comparison} with thicker solid line showing the mean and the two thinner lines
indicating the dispersion with respect to the mean.
The upper right panel of Fig.~\ref{comparison} shows the radial profile of the
anisotropy parameter
\begin{equation}	\label{beta}
	\beta=1-\frac{\sigma_t^2(r)}{2 \sigma_r^2(r)}
\end{equation}
averaged in the same way over the ten haloes, where $\sigma_t^2 = \sigma_\theta^2 + \sigma_\phi^2$
and $\sigma_{\theta,\phi}$ and $\sigma_r$ are the angular and radial velocity
dispersions with respect to the mean velocities. As we can see, the orbits
are mildly radial, with positive mean $\beta$.

The dispersions needed for the calculation of $\beta$ were approximated by $\sigma=S$ where
$S^2$ is the most natural estimator of the variance from a large sample of
$n$ velocities $v_i$
\begin{equation}    \label{app1}
	S^2 = \frac{1}{n} \sum_{i=1}^n (v_i - \overline{v})^2
\end{equation}
where $\overline{v} = (\sum_{i=1}^n v_i )/n$
is the mean of velocities in the shell. The lower left panel of Fig.~\ref{comparison}
shows the radial velocity dispersion calculated in the same way.
In the lower right panel of Fig.~\ref{comparison} we plot the fourth radial velocity
moment normalized by the variance squared, i.e. the kurtosis
\begin{equation}    \label{app2}
	K = \frac{\frac{1}{n} \sum_{i=1}^n (v_i - \overline{v})^4}{(S^2)^2}.
\end{equation}

Although the moments show a significant scatter from halo to halo and some variability
due to substructure in every halo, the overall trend in their behaviour, shown by the
mean values plotted in Fig.~\ref{comparison}, is clearly visible.
In particular, the slightly increasing and then decreasing velocity dispersion profile
agrees well with the predictions of {\L}okas \& Mamon (2001) for the NFW haloes.
The trend observed in the
behaviour of kurtosis which decreases from a value close to 3 (characteristic of a Gaussian
distribution) to values well below 3 suggests a flat-topped velocity distribution in the outer
radial bins.

\vspace{-0.25in}

\section{Comparison with predictions from the Jeans formalism}

The Jeans formalism (e.g. Binney \& Tremaine 1987) relates the velocity moments
of a gravitationally bound object to the underlying mass distribution. We summarize
here the formalism, as developed in \L okas (2002) and \L okas \& Mamon (2003).
The second $\sigma_r^2$ and fourth-order $\overline{v_r^4}$ radial velocity moments
obey the Jeans equations
\begin{eqnarray}    \label{m1}
	\frac{\rm d}{{\rm d} r}  (\nu \sigma_r^2) + \frac{2 \beta}{r} \nu
	\sigma_r^2 + \nu \frac{{\rm d} \Phi}{{\rm d} r} &=& 0   \label{jeans1} \\
 	\frac{\rm d}{{\rm d} r}  (\nu \overline{v_r^4}) + \frac{2 \beta}{r} \nu
	\overline{v_r^4} + 3 \nu \sigma_r^2 \frac{{\rm d} \Phi}{{\rm d} r} &=& 0
	\label{jeans2}
\end{eqnarray}
where $\nu$ is the 3D density distribution of the tracer population (here it is the same
as the total mass density) and $\Phi$ is
the gravitational potential, which for an NFW density distribution is
$\Phi(s)/V_v^2 = - g(c) \ln (1 + c s)/s$.
The second equation was derived assuming the distribution
function of the form $f(E,L)=f_0(E) L^{-2 \beta}$ and
the anisotropy parameter $\beta$, equation (\ref{beta}), to be constant with radius.
We will consider here $-\infty < \beta \le 1$ which covers
all interesting possibilities from radial orbits
($\beta=1$) to isotropy ($\beta=0$) and circular orbits
($\beta \rightarrow - \infty$).

The solutions to equations (\ref{jeans1})-(\ref{jeans2}) for $\beta={\rm const}$ are
\begin{eqnarray}
	\nu (r) \sigma_r^2 (r) &=& r^{-2 \beta}
	\int_r^\infty x^{2 \beta} \nu (x) \frac{{\rm d} \Phi}{{\rm d} x} \ {\rm d} x
	\label{sol1} \\
	\nu (r) \overline{v_r^4} (r) &=& 3 r^{-2 \beta}
	\int_r^\infty x^{2 \beta} \nu (x) \sigma_r^2 (x)
	\frac{{\rm d} \Phi}{{\rm d} x} \ {\rm d} x \ .
	\label{sol2}
\end{eqnarray}
After introducing expression (\ref{sol1}) into (\ref{sol2}) and inverting the order
of integration
in (\ref{sol2}) the expression for the fourth moment reduces to a single integral
\begin{equation}	\label{sol2a}
	\nu (r) \overline{v_r^4} (r) = 3 r^{-2 \beta}
	\int_r^\infty x^{2 \beta} \nu (x)
	\frac{{\rm d} \Phi}{{\rm d} x} [\Phi(x)-\Phi(r)]\ {\rm d} x \ .
\end{equation}

For the specific case of NFW-distributed dark matter particles tracing their own gravitational
potential the formulae reduce to the following expressions which can be easily
calculated numerically
\begin{eqnarray}
	\frac{\sigma_r^2 (s)}{V_v^2} &=& s^{1-2 \beta} (1+c s)^2 g(c) \label{sol1b} \\
	& \times &
	\int_s^\infty \frac{z^{2 \beta-3}}{(1+c z)^2} \left[\ln(1+c z) -
	\frac{c z}{1+ c z} \right] \ {\rm d} z       \nonumber
	\\
	\frac{\overline{v_r^4} (s)}{V_v^4} &=& 3 s^{1-2 \beta} (1+c s)^2 g^2 (c)
	\label{sol2b} \\
	& \times &
	\int_s^\infty \frac{z^{2 \beta-3}}{(1+c z)^2} \left[\ln(1+c z) -
	\frac{c z}{1+ c z} \right]     \nonumber   \\
	& \times & \left[ \frac{\ln(1+c s)}{s} - \frac{\ln (1+ c z)}{z} \right] \ {\rm d} z
	\nonumber
\end{eqnarray}
where the distances were expressed in units of the virial radius ($s=r/r_v$, $z=x/r_v$) and
$V_v$ is the circular velocity at the virial radius.
As usual, we express the fourth-order moment in terms of the dimensionless radial kurtosis
\begin{equation}	\label{kurtr}
	\kappa_r (r) = \frac{\overline{v_r^4} (r)}
	{\sigma_r^4 (r)} .
\end{equation}

\begin{figure}
\begin{center}
    \leavevmode
    \epsfxsize=8cm
    \epsfbox[55 55 270 150]{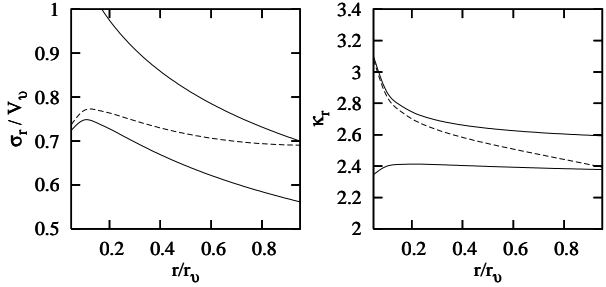}
\end{center}
    \caption{Predicted radial velocity moments for $c=7.4$ and different $\beta$.
	The left (right) panel shows the radial
	velocity dispersion $\sigma_r$ (kurtosis $\kappa_r$) for $\beta=0$ with the lower
	(upper) solid line
	and for $\beta=0.6$ with the upper (lower) solid line.
	The dashed line in each panel shows the result for $\beta$
	linearly increasing from 0 to 0.6. }
\label{predictions}
\vspace{-0.15in}
\end{figure}

Although the velocity dispersion profile depends on the concentration parameter (see
{\L}okas \& Mamon 2001), our simulated haloes all have similar concentration, so for the
comparison with the predictions from the Jeans formalism we will adopt their mean concentration
of $c=7.4$ and focus on the dependence on the anisotropy parameter $\beta$. Since the
anisotropy changes roughly between $\beta=0$ in the centre and $\beta=0.6$
at the virial radius, as shown in the upper right
panel of Fig.~\ref{comparison}, we first plot in Fig.~\ref{predictions} the velocity
moments obtained from the solutions (\ref{sol1b})-(\ref{sol2b}) for these two constant values
of $\beta$. We do not have to restrict our analysis to a constant $\beta$ however,
since the main contribution to $\sigma_r(r)$ and $\kappa_r(r)$ comes from
the region close to $r$ in the integrands given in (\ref{sol1})-(\ref{sol2})
and the moments are
actually sensitive only to a local value of $\beta$ at a given $r$. We can therefore reproduce
the moments for variable beta taking a different (constant) $\beta$ for each $r$. Assuming
that $\beta$ grows linearly between 0 and 0.6 we get results shown with a dashed line in
each of the panels in Fig.~\ref{predictions}. We see that the new solution changes smoothly
from the result for $\beta=0$ near the centre to the result for $\beta=0.6$ near the virial
radius.

In the lower panels of Fig.~\ref{comparison} we show with dashed lines
the predictions from (\ref{sol1b})-(\ref{sol2b}) for $\beta$ changing
as shown in the upper right panel. Namely, as described above, for the calculation of the
predicted moments in each bin we assume a value of $\beta$ found in the same bin
after averaging over ten haloes. We can see that the solutions of the Jeans
equations reproduce quite well the  trends in the behaviour of the radial
velocity moments, especially in the case of dispersion. Although the solutions are not
formally exact, they do provide fairly good approximations.

\begin{figure}
\begin{center}
    \leavevmode
    \epsfxsize=8cm
    \epsfbox[55 55 270 250]{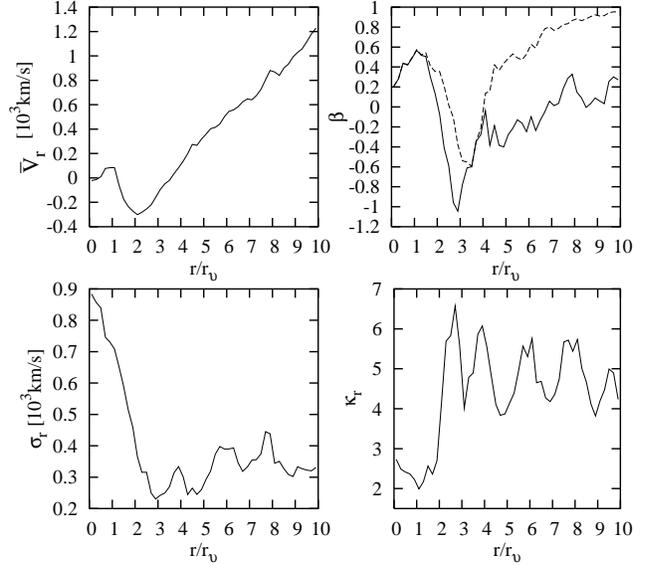}
\end{center}
    \caption{The same quantities as in Fig.~\ref{comparison} measured for a composite cluster made
	of ten haloes for distances out to 10 $r_v$. The dashed line in the upper right panel shows
	the $\beta$ parameter including the non-zero mean velocities (see text).}
\label{beyond}
\vspace{-0.15in}
\end{figure}

\begin{figure}
\begin{center}
    \leavevmode
    \epsfxsize=5.8cm
    \epsfbox[55 55 255 605]{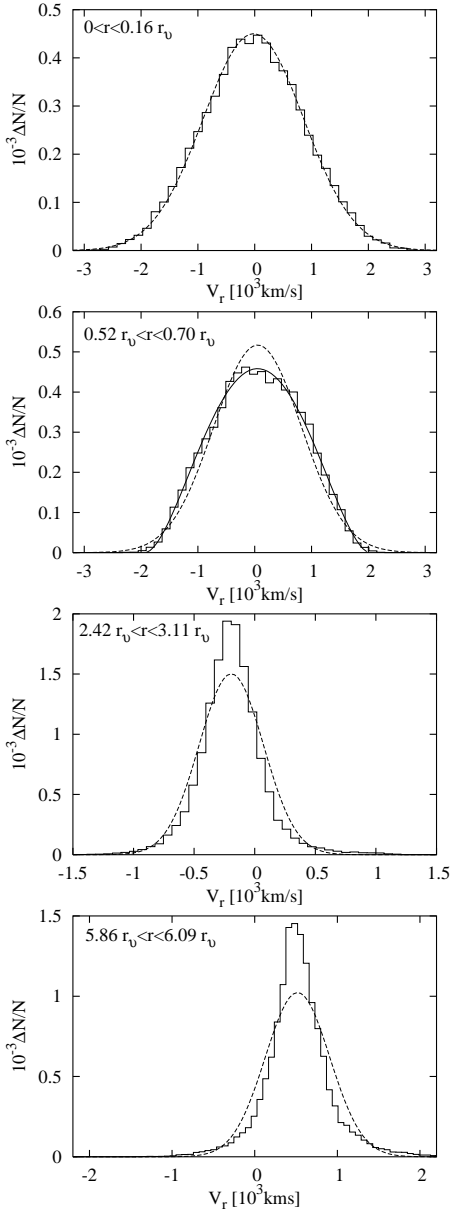}
\end{center}
    \caption{The radial velocity distribution of the composite cluster at different distances.
	The histograms
	were made for radial bins of equal number of particles. The range of particle positions
	in each bin is written in the corresponding panel.
	On top of the histograms we plot with dashed lines the normalized
	Gaussian distributions with dispersions measured in a given bin.
	The solid line in the second panel shows a best-fitting Tsallis distribution
	with $q=0.65$.}
\label{distribution}
\vspace{-0.15in}
\end{figure}

\vspace{-0.25in}

\section{Radial moments beyond the virial radius}

In this section we extend the analysis beyond the virial radius and look at the behaviour of
the moments there. Fig.~\ref{beyond} again shows the same quantities as in Fig.~\ref{comparison}
but for a composite cluster made of particles from the ten haloes with distances scaled to $r_v$
of each halo. We plot the measurements up to distances of 10 $r_v$, averaged
in radial bins of size $0.2 r_v$. The mean radial
velocity displays a characteristic behaviour: just outside the virial radius it drops
to negative velocities signifying infall, reaches the turn-around radius of the cluster at
about 3-4 $r_v$ and approaches the Hubble flow at large distances.
The velocity dispersion falls to the value characteristic of the background just outside 2 $r_v$.
The kurtosis falls down inside
$r_v$ to values below 3 and then increases sharply at larger distances and oscillates
around 5 but staying well above 3. This signifies a velocity distribution which is more peaked
in the centre than the Gaussian.
The solid line in the upper right panel of the Figure shows the anisotropy parameter defined as in
eq. (\ref{beta}). For comparison we also plot with a dashed line the parameter with the
contribution from non-zero mean velocities (i.e. $\beta'=1-\overline{v_t^2}/2 \overline{v_r^2}$ where
$\overline{v_i^2} = \sigma_i^2 + \overline{v_i}^2$).
The anisotropies drop to values below zero
outside $r_v$ and approach zero or unity at large distances. This behaviour can be understood as
due to dominance of ordered flow in the radial direction outside the virial radius: random
motions in the radial direction are smaller while the tangential ones remain the same which
results in negative~$\beta$.

\vspace{-0.25in}

\section{Summary and conclusions}

We summarize our results by plotting explicitly the radial velocity distribution of our
composite cluster in Fig.~\ref{distribution}. The histograms shown in the Figure
were made for radial bins of equal number ($ 4 \times 10^4$)
of particles and show their velocity distribution
with respect to the mean velocity in each bin. The range of particle positions
in each bin (increasing from top to bottom panel) is marked in the corresponding panel.
On top of the histograms we plot with dashed lines the
normalized Gaussian distributions with dispersions as
shown in the lower left panel of Fig.~\ref{beyond}.

The first two panels show the
distribution inside the virial radius. We can see that the distribution is Gaussian to a very
good approximation only in the very centre. In the next bin the distribution
is flat-topped with respect to a Gaussian (with the effect becoming systematically stronger
when approaching the virial radius). This behaviour is reflected in the decreasing kurtosis
profile in the lower right panel of Fig.~\ref{comparison}. The distribution shown in the
second panel turns out to be quite well fitted by Tsallis distribution
$f(v) \propto [1-(1-q) (v/v_0)^2]^{q/(1-q)}$
with entropic index $q=0.65$ and $v_0=1178$ km s$^{-1}$ which we show with a solid line
(see Lavagno et al. 1998; Hansen et al. 2004).

The third panel shows the velocity distribution in the infall region: the mean velocity
of the particles is now negative, but the shape of the distribution is also altered.
The distribution is more
peaked than a Gaussian signifying a dominance of ordered flow over random motion of particles.
The last panel illustrates the velocity distribution in the outflow region, i.e. at distances
larger than the turn-around radius. Here the distribution is also more peaked than the
Gaussian but the mean velocity of particles is positive.

We have therefore shown that, contrary to the still commonly held belief, the velocity distribution in
dark matter haloes is not Gaussian. Except for the very centre, it remains flat-topped
inside the virial radius. The transition from the virialized to the non-virialized region is marked
by a change in the velocity distribution from flat-topped to strongly
peaked with respect to the Gaussian.
We have demonstrated that the velocity moments in the virialized region are quite well reproduced by the
solutions of the Jeans equations.

The Gaussian shape of the velocity distribution in the centre is expected since this kind
of distribution is characteristic of structures with isothermal density profiles ($r^{-2}$) and
isotropic orbits (Binney \& Tremaine 1987; Hansen et al. 2004). Hansen et al. also show that
for isotropic orbits steepening of the density profile results in flattening of the velocity
distribution, as we observe in our haloes despite the fact that their orbits are mildly radial.
It would be very interesting to fully understand the relation between the shape
of the distribution and the density profile and anisotropy of particle orbits.

\vspace{-0.25in}

\section*{Acknowledgements}

We wish to thank M. Chodorowski and an anonymous referee for their comments.
Computer simulations presented in this paper were performed at the
Leibnizrechenzentrum (LRZ) in Munich.
RW acknowledges the summer student program at Copernicus Center.
RW and E{\L} are grateful for the hospitality of Astrophysikalisches
Institut Potsdam where part of this work was done.
This research was partially supported by the
Polish Ministry of Scientific Research and Information Technology
under grant 1P03D02726 as well as
the Jumelage program Astronomie France Pologne of CNRS/PAN.

\end{document}